\documentclass[
    ,final            
    ,numberedheadings 
  ]
  {aipproc}

\layoutstyle{6x9}

\def \la{\mathrel{\mathchoice   {\vcenter{\offinterlineskip\halign{\hfil
$\displaystyle##$\hfil\cr<\cr\sim\cr}}}
{\vcenter{\offinterlineskip\halign{\hfil$\textstyle##$\hfil\cr
<\cr\sim\cr}}}
{\vcenter{\offinterlineskip\halign{\hfil$\scriptstyle##$\hfil\cr
<\cr\sim\cr}}}
{\vcenter{\offinterlineskip\halign{\hfil$\scriptscriptstyle##$\hfil\cr
<\cr\sim\cr}}}}}

\def \ga{\mathrel{\mathchoice   {\vcenter{\offinterlineskip\halign{\hfil
$\displaystyle##$\hfil\cr>\cr\sim\cr}}}
{\vcenter{\offinterlineskip\halign{\hfil$\textstyle##$\hfil\cr
>\cr\sim\cr}}}
{\vcenter{\offinterlineskip\halign{\hfil$\scriptstyle##$\hfil\cr
>\cr\sim\cr}}}
{\vcenter{\offinterlineskip\halign{\hfil$\scriptscriptstyle##$\hfil\cr
>\cr\sim\cr}}}}}

\begin{document}

\title{Stellar Populations\\ \medskip in the Local Group of Galaxies}


\author{Eva K.\ Grebel}{
  address={Astronomical Institute of the University of Basel, 
           Venusstrasse 7, CH-4102 Binningen, Switzerland}
}

\begin{abstract}
 The characteristics and properties of the stellar populations and 
 evolutionary histories of Local Group galaxies are summarized and 
 compared to predictions of cosmological models.  No clear signature
 of the re-ionization epoch is observed; in particular, there is no
 cessation of star formation activity in low-mass dwarf 
 galaxies at the end of re-ionization.  Arguments against
 the morphological transformation of dwarf irregular into dwarf
 spheroidal galaxies are derived from their pronounced evolutionary
 differences at early epochs as evidenced by the offset in the
 metallicity-luminosity relation between gas-rich and gas-poor dwarfs.
 While there is increasing evidence for past and ongoing accretion 
 events the overall importance of dwarf galaxies as building blocks remains
 unclear considering their differences in modes of star formation
 and detailed chemistry. 
\end{abstract}

\maketitle


\section{The Local Group}

The Local Group of galaxies is dominated by two massive spirals, M31 and 
the Milky Way.  The Local Group (LG) is our immediate cosmic neighborhood, 
a sparse galaxy group with a zero-velocity radius of 1 Mpc (Karachentsev 
et al.\ 2002a).  Two dominant galaxies also characterize a number of other
nearby galaxy groups, and zero-velocity radii of 1 to 1.2 Mpc are typical
for these poor groups.  Thirty-eight galaxies are currently known to be located 
within the LG's radius, but the galaxy census of the LG is still incomplete.  
While all of its luminous members have been detected, new faint member 
candidates with very low surface brightnesses are still being discovered 
(e.g., Zucker et al.\ 2004a).  In spite of its small size and its small 
number of galaxies the LG plays a prominent role in astrophysical research.  
For reviews of the LG, see Grebel (1997, 1999, 2000, 2001), Mateo (1998),
and van den Bergh (1999, 2000).  In the present contribution, I will mainly
concentrate on the dwarf galaxies of the LG. 

The proximity of the galaxies of the LG permits us to study them at the
highest possible resolution, enabling us to resolve these galaxies into
individual stars down to very faint magnitudes.  For example, with the 
Hubble Space Telescope, point sources down to an apparent magnitude of 
$\approx$ 30 have been detected in M31 (Brown et al.\ 2003), 
corresponding to the magnitudes
of stars well below the Population II main-sequence turn-off at that distance.
To date, the LG is the
only location where we can observe the lowest stellar masses, measure the
oldest stellar ages, determine metallicities and element abundance ratios
of individual stars of high ages, and measure detailed stellar
kinematics.  In other words, the LG is unique in affording the
highest level of detail and accuracy for stellar population studies.  Much
of this research has only become possible in recent years thanks to the
superior angular resolution and sensitivity of the Hubble Space Telescope
(HST),
and thanks to the advent of powerful ground-based optical/near infrared 8 
to 10m class telescopes, which provide sufficient sensitivity for medium 
and high-resolution stellar spectroscopy of individual stars in nearby 
galaxies.  It is now becoming possible to uncover the evolutionary histories
of these galaxies in unprecedented detail. 

Furthermore, the LG contains a variety of different galaxy types.    
While massive early-type galaxies are absent, the LG still provides
us with galaxies covering five orders of magnitude in galaxy masses,
a variety of different morphological types, very different star formation
histories, and a range of different dominant ages and metallicities.  
Within the small volume of the LG, galaxies occur in environments 
ranging from immediate proximity to massive galaxies to fairly isolated 
locations without nearby neighbors.  Hence we can expect to eventually
obtain first-hand measurements of the impact of environment on galaxy
evolution, at least on small scales.

All in all, by studying the stellar populations of galaxies in the LG
and by combining them with detailed knowledge of the other components of
these galaxies such as gas, dust, and non-luminous matter we can learn
about galaxy properties at the highest possible level of detail.  The
LG permits us to study stellar evolution at a range of different 
metallicities and ages not afforded in that combination in the Milky Way;
e.g., the evolution of young, massive stars at low metallicities.  
Moreover, the LG permits us to carry out tests of cosmological galaxy 
evolution theories, for instance with respect to time scales and number
and magnitude of accretion events.  Hence here we can conduct ``near-field
cosmology'' (see Freeman \& Bland-Hawthorn 2002).  The LG is particularly
well-suited to learn more about the formation and evolution of disk
galaxies, whose theoretical foundations are still poorly understood.
Understanding all of these processes in an environment where we can 
learn about them in detail is the precondition for understanding distant,
unresolved galaxies.  

\subsection{The galaxy content of the Local Group}

We currently know of 38 galaxies within the zero-velocity surface of the
Local Group.  Whether the outermost galaxies are indeed gravitationally
bound to the LG remains unclear.  Indeed, reliable orbits are not yet 
known for any of the LG galaxies, and proper motion measurements exist
only for the closest Milky Way companions.  The uncertainties of these
measurements are still large, but this situation should change once data
from the planned astrometric satellite missions -- Gaia (an ESA mission
that will scan repeatedly the entire sky) and the Space Interferometry
Mission (SIM, a NASA mission that will perform pointed observations with
even higher accuracy) -- become available.  These missions will probably
not start before 2011.  

As mentioned before, the LG galaxy census is still incomplete.  Apart
from missing orbital information, galaxies of very faint surface brightness
are still being detected as sky surveys become more sensitive or as data
mining techniques are being improved.  During the past six years, this led 
to the discovery of one fairly isolated new dwarf spheroidal (dSph) galaxy
in the Local Group (Cetus; Whiting, Irwin, \& Hau 1999), and to the 
discovery of four new dSph companions of M31 (Armandroff,  Davies, \& Jacoby
1998; Armandroff, Jacoby, \& Davies 1999;
Karachentsev \& Karachentseva 1999, Grebel \& Guhathakurta 1999, 
Zucker et al.\ 2004a, Harbeck et al.\ 2005).  
The conclusion of data mining existing photographic all-sky surveys
(Karachentsev et al. 2000; Whiting, Hau, \& Irwin 2002) makes the existence
of additional faint, Draco- or Ursa-Minor-like dSphs seem unlikely unless
they are hidden behind regions of high Galactic extinction, particularly
in the zone of avoidance.  On the other hand, the discovery of And IX
(Zucker et al.\ 2004a) with an unusually low surface brightness of only
$\mu_{V,0} \sim 26.8$ mag arcsec$^{-2}$ indicates that additional very sparse,
very faint dwarfs may yet have to be discovered.  At present, though,
there is little reason to believe that there may still be hundreds of
yet to be detected faint objects present in the LG.   

Very faint, sparse, low-surface-brightness objects like And IX, which
appear to be dominated by ancient, metal-poor stars (Harbeck et al.\ 2005)
are of immense interest from the point of view of galaxy evolution.  Do
they show signatures of cosmic re-ionization squelching?  Do they contain
the purest, most metal-poor, most ancient populations  known?  Or do they
show evidence for population gradients in their old population as were
found in a number of other ancient dSphs (Harbeck et al.\ 2001)?  Do 
these objects show large metallicity spreads as measured in other seemingly
purely old dSphs (Shetrone, C\^ote, \& Sargent 2001), which in turn 
indicate extended episodes of star formation (Ikuta \& Arimoto 2002)? 
How did these objects survive?  Are they entirely dominated by dark matter?  
Are they regular dSph galaxies that simply are fainter and sparser than the
previously known members of this galaxy class?  Yet
from a cosmological point of view, objects of this type may appear to
be of little interest since they may not significantly contribute toward
solving the ``missing satellite problem'' (Klypin et al.\
1999; Moore et al.\ 1999) unless detected in vast numbers. 

Not only low surface brightness, scarcity of stars, high foreground
extinction, and unknown orbits for distant LG member candidates contribute 
to an incomplete LG galaxy census, but also accretion events.  We still
know little about the actual number of accretion events, about the times
when these events occurred, and about the nature of the accreted galaxies
(see also the Section 1.4 and the contribution of Majewski in these
proceedings).  The number of 38 probable LG members comprises all galaxies
listed in Grebel, Gallagher, \& Harbeck (2003) including the Sagittarius
dwarf galaxy, which is currently being accreted by the Milky Way.  In
addition, it includes the giant stream around M31 (Ibata et al.\ 2001),
which is very likely the remnant of a
disrupted galaxy, and the newly discovered M31 satellite And IX (Zucker
et al.\ 2004a).  {\em Not}
included are other possible streams and accretion remnants such as the
Monoceros feature (Newberg et al.\ 2002) and Canis Major (Martin et al.\ 2004a)
since their nature as dwarf galaxy remnants remains disputed (see 
Section 1.4).  Depending on the nature of these and other stellar overdensities
and kinematically identified features in the Milky Way and in M31, the
LG census may need to be increased by at least
up to six or more galaxies if one
wishes to account for still measurable accretion events.  If these 
features are confirmed as minor mergers, then it becomes a matter of
definition whether their progenitors should be included in the current
LG member census.  

\subsection{Galaxy types in the Local Group}

Approximately 90\% of the LG's luminosity is contributed by its three
spiral galaxies, but the vast majority of the LG galaxy
population are satellite and dwarf galaxies.  The boundaries between
larger galaxies and dwarf galaxies are poorly defined.  We adopt here
the usual convention of calling every galaxy with an absolute magnitude
of $M_V > -18$ mag a dwarf galaxy.  This simple magnitude criterion 
then leaves us with 34 dwarf galaxies (though admittedly the absolute
magnitude of the progenitor of M31's ``great stream'' is yet unknown, and
that of the Sagittarius dSph is poorly constrained).  Apart from the
three spirals, only the Large Magellanic Cloud is more luminous than
the above luminosity threshold.  Yet M32 is often not considered a
dwarf galaxy despite its lower luminosity since it  exhibits similar 
structural properties as giant ellipticals.

Most of the LG galaxies belong to one of four basic classes:  Spirals,
dwarf irregular galaxies, dwarf elliptical galaxies, and dSphs.  In 
the following paragraphs the main characteristics of the dwarf galaxy
classes are summarized following Grebel (2001):

{\bf Dwarf irregular galaxies} (dIrrs) are gas-rich galaxies with an
irregular optical appearance usually dominated by scattered
H\,{\sc ii} regions.  They typically have V-band surface brightnesses of
$\mu_V \la 23$ mag arcsec$^{-2}$, H\,{\sc i} masses of
$M_{\rm HI} \la 10^9$ M$_{\odot}$, and total masses of
$M_{\rm tot} \la 10^{10}$ M$_{\odot}$.  The stellar populations of dIrrs
range from ancient stars with ages $>10$ Gyr to ongoing star formation.
Star formation appears to have proceeded largely continuously, although
amplitude variations in the intensity or rate of star formation may reach 
a factor of three.  The H\,{\sc i} distribution is usually clumpy and
more extended than even the oldest stellar populations.
When considering only the distribution of older stellar populations (i.e., of
red giants and red clump stars, which trace populations older than $\sim 1$
Gyr), irregulars and dIrrs exhibit a highly regular distribution
shaped like an (elongated) disk (LMC; van der Marel 2001)
or like a spheroid (example: Small Magellanic Cloud (SMC); Zaritsky et al.\
2000). 
In low-mass dIrrs gas and stars may exhibit distinct spatial distributions
and different kinematic properties.  Metallicities tend to increase with
decreasing age in the more massive dIrrs, indicative of enrichment
and of an age-metallicity relation (see also Fig.\ 3 in Grebel 2004).
For low-mass dIrrs detailed measurements are still lacking.
Solid body rotation is common among the more massive dIrrs,
whereas low-mass dIrrs seem to be dominated by random motions without
evidence of rotation.
DIrrs are found in galaxy clusters, groups and in the field.

{\bf Dwarf elliptical galaxies} (dEs) are spherical or elliptical in
appearance, tend to be found near massive galaxies (in the Local
Group they are all companions of M31),
usually have little or no detectable gas, and tend not to be
rotationally supported.  Note that examples of rotating dEs beyond
the Local Group have been found (e.g., Pedraz et al.\ 2002).
DEs are compact galaxies with high central
stellar densities.  They are typically fainter than M$_V = -17$ mag,
have $\mu_V \la 21$ mag arcsec$^{-2}$, $M_{\rm HI}
\la 10^8$ M$_{\odot}$, and $M_{tot} \la 10^9$ M$_{\odot}$.
DEs may contain conspicuous nuclei (nucleated dEs or dE,N) that
may contribute up to 20\% of the galaxy's light.  The fraction of dE,N
is higher among the more luminous dEs.  An example for a non-nucleated
dE in the Local Group is NGC\,185, whereas NGC\,205 is a dE,N.
S\'ersic's (1968) generalization of a de Vaucouleurs $r^{1/4}$ law
and exponential profiles describe the surface
density profiles of nucleated and non-nucleated dEs and dSphs best
(Jerjen et al.\ 2000).  DEs typically contain old and intermediate-age
populations (i.e., populations older than 10 Gyr and populations in
an age range of $\sim 2$ to 10 Gyr), but the fractions of these 
populations vary, and even present-day star formation may be observed.

{\bf Dwarf spheroidal galaxies} (dSphs) are diffuse, gas-deficient,
low-surface-brightness
dwarfs with very little central concentration.  They are not always
distinguished from dEs in the literature.  DSphs are characterized by
M$_V \ga -14$ mag, $\mu_V \ga 22$ mag arcsec$^{-2}$,
$M_{\rm HI} \la 10^5$ M$_{\odot}$, and $M_{tot} \sim 10^7$ M$_{\odot}$.
They include the optically faintest galaxies known.  Their stellar
populations tend to be either almost purely old or a mix of old and
intermediate-age populations.  The luminosity functions (and by inference
the mass functions) of dSphs have been found to be ``normal'' and in
excellent agreement with those of Galactic globular clusters (Grillmair
et al.\ 1998; Wyse et al.\ 2002).  DSphs are usually found
in close proximity of massive galaxies (counterexamples in the LG:
Cetus, Tucana) and are generally not supported by rotation.  Since most 
published measurements concentrated on the central regions of dSphs, the 
possibility of slowly rotating dSphs cannot be excluded.  The central 
velocity dispersions of dSphs indicate the presence of a significant dark
component when virial equilibrium is assumed.  However, not all dSphs are
in virial equilibrium (e.g., Ursa Minor shows indications of being tidally
distorted by the Milky Way; for instance, Palma et al.\ 2003).
Interestingly, the radial velocity dispersion profiles of dSphs show a 
marked drop at large radii (Wilkinson et al.\ 2004).  Its interpretation
and the apparent existence of a kinematically cold stellar population at
the outermost radii is not yet understood. 
The metallicity--luminosity relations of dSphs and dIrrs show the usual
trend of increasing metallicity with increasing galaxy luminosity, but the
relations are offset from each other: DSphs have higher mean stellar
metallicities at a given optical luminosity (Grebel et al.\ 2003 and
Section 3.1).

\subsection{Morphological segregation}

\begin{figure}
  \includegraphics[height=.4\textheight]{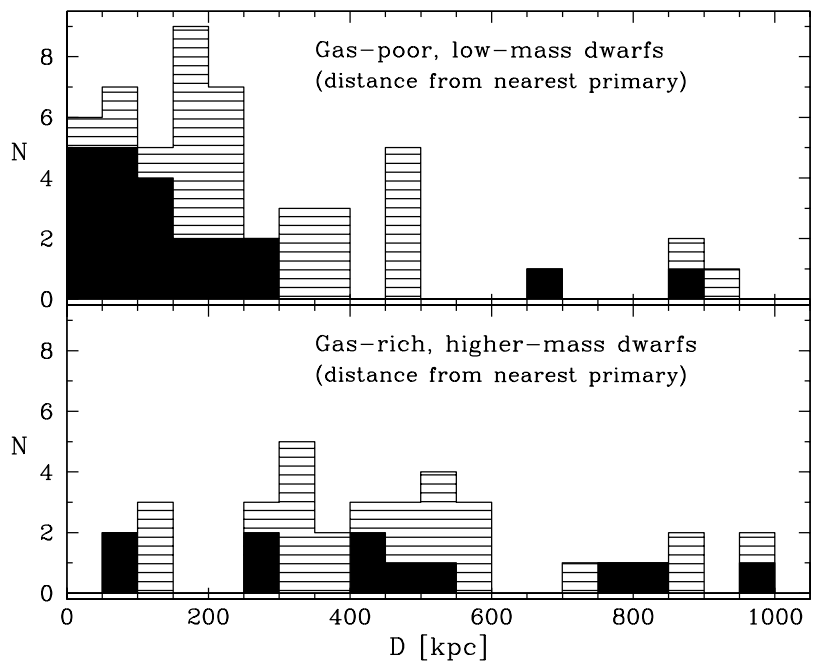}
  \caption{Morphological segregation in the number distribution N
of different types of galaxies in the Local Group (solid histograms)
and in the M81 and Centaurus\,A groups (hashed histograms) as a 
function of distance D to the closest massive primary (updated version
of Fig.\ 1 in Grebel 2004).}
\end{figure}

As already indicated, the distribution of the different types of galaxies
within the LG is not random, but is determined by the location of the two 
most massive spiral galaxies in the LG, M31 and the Milky Way.  A sketch
of the three-dimensional distribution of galaxies within the LG is shown
in Grebel (1999; in Fig.\ 3).  In Fig.\ 1 the distribution of gas-poor,
low-mass dwarfs and of gas-rich, higher-mass dwarfs is shown.  The gas-poor
dwarfs -- primarily dSphs -- are strongly clustered around the closest
massive spiral, whereas the gas-rich dwarfs (dIrrs) exhibit less of a 
tendency toward close concentration around massive galaxies and are more 
widely distributed.  The differing distribution of different morphological
galaxy types is also known as morphological segregation and is observed in
nearby galaxy groups and galaxy clusters as well.  

Similarly, the H\,{\sc i}
mass of dwarf galaxies tends to increase with increasing distance to a 
massive primary (Grebel et al.\ 2003; their Fig.\ 3).  These trends 
indicate that environment may have a significant impact on the evolution
of low-mass galaxies.  Indeed, it is tempting to speculate that these 
trends are the result of morphological transformations due to the 
influence of massive galaxies, e.g., via tidal or ram pressure stripping
(see Mayer et al.\ 2001 for simulations).
One vital ingredient to verify this hypothesis   is once again information
about the orbits of the companion galaxies. 

\subsection{Direct evidence for harassment and accretion}

If accretion is the primary mechanism for the growth and evolution of
massive galaxies as predicted by cosmological models, then we should
be able to find evidence for these processes in our immediate neighborhood.
(1) The study of the structural properties of nearby galaxies can reveal 
whether external tidal forces are distorting them.  (2) The detection of
extratidal stars and streams around and within massive galaxies is
evidence for ongoing harassment and accretion events.  (3) The stellar
content, population properties, and chemistry of nearby galaxies allow
us to constrain to what extent these kinds of objects could have 
contributed as building blocks to more massive galaxies.   

The clearest evidence for ongoing accretion are the extended tidal
stream of the Sagittarius dSph galaxy (Ibata, Gilmore, \& Irwin 1994),
which has now been traced around the entire Milky Way using M giants
identified in the Two Micron All-Sky Survey (2MASS) (Majewski et al.\ 
2003), and the giant
stream of metal-rich giants in the halo of M31 (Ibata et al.\ 2001;
Ferguson et al.\ 2002).  Additional stellar overdensities have been
detected in the Milky Way using various photometric data sets including
2MASS and the Sloan Digital Sky Survey (SDSS):  The Monoceros feature
(Newberg et al.\ 2002; Yanny et al.\ 2003), which may be a tidal tail 
connected with the Canis Major overdensity (Martin et al.\ 2004a).  The
interpretation of Canis Major is disputed; suggestions include that it
is part of the Galactic warp or flare (Momany et al.\ 2004) or indeed
the center of another possibly disrupted dSph within the Milky Way 
(Martin et al.\ 2004a, 2004b).  Additional Galactic stellar 
overdensities have been identified (for instance, Triangulum-Andromeda;
Rocha-Pinto et al.\ 2004, which may be part of the tidal tail of a 
more distant disrupted dwarf).  Other suggestions of evidence of dwarf
galaxy accretion are based on the identification of moving groups and
radial velocity surveys (e.g., Gilmore, Wyse, \& Norris 2002). 
However, not only disrupted dwarf galaxies, but even globular clusters
in advanced stages of accretion may produce extended stellar tidal
tails (Odenkirchen et al.\ 2001; 2003).  These tidal features are valuable
also as tracers of the Galactic potential.  The most luminous and most
massive Galactic globular cluster, $\omega$ Centauri, contains a range
of different ages and a large metallicity spread.  A popular explanation
for its unusual properties is that $\omega$ Cen may be the core of an
accreted dwarf galaxy (see van Leeuwen, Hughes, \& Piotto 2002 for 
details).  --- Possible additional tidal
features in and near M31 have been reported by Morrison et al.\ (2003)
and Zucker et al.\ (2004b). 
 
The Magellanic Clouds and the Milky Way are interacting with each other
as evidenced by, e.g.,  the gaseous
Magellanic Stream and the Magellanic Bridge (e.g., Br\"uns et al.\
2005), although the interpretation of these features as being primarily
due to tidal (e.g., Putman et al.\ 1998) or ram pressure effects
(Mastropietro et al.\ 2005) remains controversial.  
The twisted isophotes of the M31
companions M32 and NGC\,205 may be caused by tidal interaction with M31
(Choir, Guhathakurta, \& Johnston 2002).  The nearby Galactic
dSph satellite Ursa Minor shows a distorted, S-shaped surface density
profile possibly caused by tidal interaction with the Milky Way
(Palma et al.\ 2003).  On the other hand, the drop
in the velocity dispersion profiles of Draco and Ursa Minor at large radii
(Wilkinson et al.\ 2004) and the lack of a large depth extent of Draco
(Klessen, Grebel, \& Harbeck 2003) would seem to argue against ongoing
tidal {\em disruption}.   

\section{The Earliest Epoch of Star Formation}

According to hierarchical structure formation scenarios, low-mass systems
should have been the first sites of star formation in the Universe.  The
first star formation events may have occurred as early as at a redshift
of z $\sim$ 30, an epoch unobservable for us with our present tools.  
Larger systems should then have formed through hierarchical merging of
smaller systems, leading to the idea of dwarf galaxies as building blocks
of more massive galaxies.  Hence one important test to carry out is to
compare the properties of the old populations in dwarf galaxies with those
in massive galaxies to investigate how similar or dissimilar they are.  If
today's dwarf galaxies are the few survivors of a once much more numerous,
since accreted low-mass galaxy population, and if dwarf galaxy accretion
is the primary process governing the formation of more massive galaxies,
then detailed studies of their old stellar populations should reveal very
similar properties.

A number of cosmological models predict that 
cosmic re-ionization will squelch star formation in low-mass
substructures, and that galaxies less massive than $10^8$ to $10^9$ M$_{\odot}$
will lose their star-forming material through photoevaporation during
re-ionization (e.g., Ferraro \& Tolstoy 2000; Dekel \& Woo 2003;
Susa \& Umemura 2004).  As a consequence, low-mass galaxies must form 
their stars prior to re-ionization (e.g., Tassis et al.\ 2003) and need 
to contain ancient populations.  Furthermore, 
one would expect a sharp drop and indeed a complete cessation of star-forming
activity after re-ionization is complete.  A third consequence is that
the oldest populations in high-mass galaxies must be either as old as 
those in low-mass galaxies, or younger.  

\begin{figure}
  \includegraphics[height=.4\textheight]{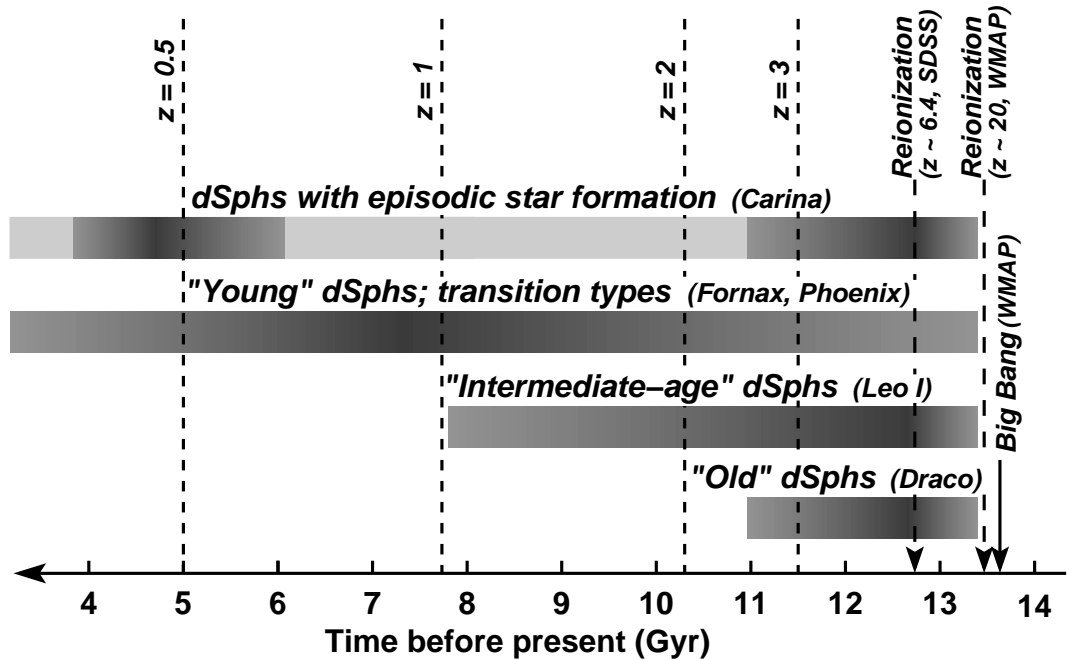}
  \caption{Bar diagram indicating the approximate duration of star formation
episodes in low-mass galaxies ($\sim 10^7$ M$_{\odot}$).  The approximate
beginning and end of the re-ionization epoch are indicated (based on results
from WMAP and from the Sloan Digital Sky Survey).  The predicted
cessation of star formation in low-mass galaxies, esp.\ in dSphs,
is not observed.  For more details, see also Grebel \& Gallagher (2004).
}
\end{figure}

These are testable predictions that can be investigated by exploiting the
fossil stellar record in nearby galaxies.  The LG is an ideal target
since here the oldest populations are resolved and can be accessed with
HST and increasingly also with large ground-based telescopes operating
with high angular resolution.  This requires age dating of old populations.
The most accurate ages can be obtained for resolved stellar populations.
For old populations the most age-sensitive feature is the old main-sequence
turn-off (MSTO), where via differential age-dating techniques internal 
accuracies of less than 1 Gyr can be obtained.  Comparison objects are 
usually ancient Galactic globular clusters of the same metallicity as the 
target population.  Absolute ages are more difficult to determine since 
here one needs to rely on isochrones, which makes the resulting ages 
model-dependent.  But also differential techniques have a number of 
drawbacks:  They require very deep, 
high-quality photometry reaching at least 2 mag below the MSTO, which 
is a challenge in more distant galaxies.  They require stellar populations 
sufficiently numerous to produce a measurable MSTO, which necessarily 
limits us to Population II stars (note that to date not a single Population
III star candidate has ever been detected beyond the Milky Way).  Ideally,
one wishes to compare populations with the same [$\alpha$/Fe] ratio. 
(Establishing to what extent this condition is met and what the consequences
are will be an important 
area for large ground-based telescopes in the coming years.)  Other effects
such as diffusion may further affect relative measurements; see Chaboyer's
contribution in these proceedings for details. 

For a more detailed discussion of the results of differential age dating
as applied to ancient Milky Way populations and nearby dwarfs I refer to
Grebel (2000) and Grebel \& Gallagher (2004).  Here only the results 
shall be summarized, which are based on
Galactic and extragalactic globular clusters and field populations with 
MSTO photometry:  {\bf (1) Old populations are ubiquitous, but their fractions
vary.  There is not a single dwarf galaxy studied in sufficient detail
{\em without} an old population.
(2) There is evidence for a common epoch of star formation.  The ancient 
Population II in the Milky Way, the LMC, and the Galactic dSphs are coeval 
within $\sim$ 1 Gyr} (which is consistent with the building block scenario).
{\bf (3) In contrast to predictions from cold dark matter models, no cessation
of star formation activity is observed during or after re-ionization.
(4) Instead, even the least massive galaxies known, the dSphs, show evidence
for star formation extending over many Gyr.}  This holds even for those
dSphs entirely dominated by very old populations -- their metallicity
spread requires enrichment due to star formation extending over several
Gyr.

It is certainly correct to point out that the accuracy of age determinations
does not permit us yet to confidently state when exactly early star formation
began and ended in dSphs.  {\em Nevertheless, the presence of ancient 
populations
in dSphs is a fact}.  The {\em large metallicity spread} in old dSphs has been 
proven spectroscopically.  Furthermore, {\em the presence of intermediate-age
populations in addition to ancient stars in many though not all dSphs cannot
be refuted}.  In only one of these dSphs (Carina) clearly episodic star
formation with well-separated ``bursts'' has been observed (Smecker-Hane
et al.\ 1994); in all others star formation appears to have
proceeded fairly continuously.  The characteristics of the star formation and 
enrichment histories of dSphs contradict the cosmological predictions
mentioned earlier, particularly the suggestion of complete photoevaporation
of baryonic material not yet turned into stars.  

There are several possible ways out of this quandary.  Either the above
quoted cosmological models are wrong, or do not take other effects that
might prevent photoevaporation into account, or the galaxies we observe
today as dSphs once were substantially more massive.  In case if the 
latter, today's dSphs would once have needed to be at least a factor of
100 more massive.  An oversimplified estimate shows that at the times when
dSphs were forming stars, they should have had at least 10 times more
baryonic mass if we assume a star formation efficiency of $10\%$. Their
present-day baryonic masses are of the order of $10^6$ M$_{\odot}$, so an
increase by one order of magnitude would not yet change the overall 
estimated masses of dSphs drastically, but it would considerably alter
their mass-to-light ratios.  More detailed calculations and more 
observational data are needed to investigate whether substantially
more massive dSph progenitors are plausible.  The differences in the
metallicity-luminosity relation of dSphs and dIrrs seem to rule out
dIrrs as progenitors of dSphs (Grebel et al.\ 2003). 

\section{Star Formation Histories and Metallicities}

Much has been said already about star formation histories in the 
preceding sections.  In the past years, considerable effort has
been spent on uncovering star formation histories of nearby dwarf
galaxies by photometric means, i.e., via color-magnitude diagrams
of the resolved stellar populations of the stars in these galaxies. 
Basically,
very detailed star formation histories are derived photometrically
through synthetic color-magnitude diagrams (CMDs) generated from isochrones
assuming an initial mass function and a variety of different 
time-dependent star formation and chemical evolution
histories.  The synthetic CMDs are compared to the observed CMDs
using various statistical techniques until the best match is 
found (e.g., Tosi et al.\ 1991; Gallart et al.\ 1996; Tolstoy \& Saha 1996;
Dolphin 1997; Holtzman et al.\ 1997).  These methods necessarily have 
to make a number of assumptions (such as which initial mass function 
and which binary fraction to adopt), rely strongly on the chosen set of 
isochrones, and are plagued by the age-metallicity degeneracy, but yield  
surprisingly similar results in spite of the different approaches of
different groups (see contributions in Lejeune \& Fernandes 2002).

The information gained from deep CMDs can be complemented by information
provided by individual age tracer stars such as Wolf-Rayet stars, carbon
stars, RR Lyrae stars, etc.\ (see Grebel 1997, 1999) to yield a more 
complete picture of the star formation histories of the galaxies in 
question.  In particular, such tracer populations are valuable when their 
specific host populations are too sparse to appear prominently in the 
CMDs.  Furthermore, star formation histories vary as a function of
position with the oldest populations being typically the most extended
ones (e.g., Harbeck et al.\ 2001), hence ideally one wishes to cover
the entire target galaxy.

Our current knowledge about the modes of star formation in LG
dwarf galaxies can be summarized as follows:  Irregular and dIrr
galaxies are characterized by largely continuous star formation with
amplitude variations of factors 2--3, governed mainly by internal,
local processes.  In the more massive irregulars with large gas
reservoirs, star formation is likely to proceed for another Hubble
time (Hunter 1997).  Low-mass so-called dIrr/dSph transition-type
galaxies are characterized by currently very low star formation rates
and may eventually turn into quiescent dSphs.
A detailed review of the evolutionary histories
of these galaxies, large-scale and small-scale star formation properties,
and their chemical evolution is given in Grebel (2004).  

DE and dSph galaxies also tend to have continuous star formation rates,
but with decreasing intensity.  Some had their peak activity at very
early ages, others several Gyr ago.  Star formation tends to be 
longest-lasting in the centers of these galaxies, and in a number
of cases age (and possibly metallicity) gradients are observed (Harbeck
et al.\ 2001).  
Repetitive or episodic star formation, as mentioned earlier, has so
far only found in Carina.  For more detailed reviews, see Grebel
(2000, 2001).  A major puzzle is the surprising lack of gas in dSph
galaxies, where (with the exception of Sculptor; Bouchard, Carignan,
\& Mashchenko 2003) only upper limits for neutral and ionized gas
could be determined (Gallagher et al.\ 2003; Grebel et al.\ 2003,
and references therein).  Interestingly, the limits for neutral
hydrogen lie well below the amounts expected from gas loss from
the old red giants in the dSphs.  Fornax, the only dSph galaxy
with star formation as recent as $\sim 200$ Myr ago (Grebel \& Stetson
1999), surprisingly also appears
to be devoid of gas.  {\bf It is neither understood what caused the gas
loss to begin with nor how it is sustained}, especially in the distant,
isolated dSphs like Cetus and Tucana.

When illustrating star formation histories via population boxes
(e.g., Grebel 1997, 1999, 2000), it becomes quickly obvious that no 
two dwarf galaxies -- not even low-mass dSphs -- share the same
star formation history.  Each galaxy needs to be considered as an
individual with regard to the time scales of its star formation and
the degree of enrichment.  A trend of increasing intermediate-age
population fractions with increasing distance from the Milky Way
among dSphs 
was first noted by van den Bergh (1994), who attributed it to the 
possible environmental impact of the Milky Way.  Star-forming material
might have been removed earlier on from the closer Galactic 
companions via ram-pressure or tidal stripping, supernova-driven winds 
or the high UV-flux from the proto-Milky Way.  
On the other hand, if environment is primarily responsible for gas-poor
dSphs, then the existence of the isolated dSphs Cetus
and Tucana is difficult to understand.  Again, knowledge of the orbits
of these galaxies would be very helpful. 

We may then turn to the M31 dSph companions, which cover a similar
range of distances as the Galactic ones.  {\bf Interestingly, these dSphs
do {\em not} show any clear correlation between star formation history
and present-day distance to M31}.  Their lack of a pronounced red clump
and of substantial number carbon stars shows a lack of any prominent
intermediate-age population regardless of their distance (Harbeck
et al. 2001, 2004, 2005).  Going one step further, the comparison of
the stellar populations in M31's dSphs and of M31's halo shows that
the dSphs cannot have been primary building blocks of M31's halo since 
it was found to be dominated by intermediate-age, comparatively metal-rich
populations (Brown et al.\ 2003).   (An old, metal-poor halo population,
however, is present as well; Brown et al.\ 2004). 

\subsection{Mean metallicities and the metallicity-luminosity relation}

Galaxies generally obey a metallicity-luminosity relation such that more
luminous (and potentially more massive) galaxies are more metal-rich
than faint galaxies with presumably more shallow potential wells.
DIrrs and dSphs differ in their metallicity-luminosity relations
(e.g., Binggeli 1994).  Originally this was discovered when comparing
nebular H\,{\sc ii} region oxygen abundances of dIrrs with stellar
metallicity estimates of gas-poor dwarfs such as dSphs.  However, in
making this comparison, very different populations are compared, and
very different metallicity tracers are used.  Nebular oxygen abundances
cannot easily be translated into mean stellar metallicities and vice
versa.  Mean stellar metallicities usually rely on the measurement of
iron or of an element that has been found to be an excellent tracer of iron
in certain populations, such as the near-infrared Ca\,{\sc ii} triplet
(Armandroff \& Da Costa 1991; Rutledge, Hesser, \& Stetson 1997; Cole 
et al.\ 2004).  These mean stellar metallicities are usually quoted
as [Fe/H] or [Me/H] (to indicate that primarily
{\em metallicity} as opposed to solely {\em Fe} is meant).  

In order to avoid the uncertain O to Fe conversion, Richer, McCall, \& 
Stasinska (1998) published a metallicity-luminosity relation based 
entirely on nebular O measurements:  H\,{\sc ii} region abundances in
dIrrs and planetary nebula (PN) abundances in dEs and dSphs.  The
offset between the resulting relations remained; however, one still
compares different populations with each other.   H\,{\sc ii} region
abundances trace the present-day abundances of the most recently
formed population of stars.  PNe trace primarily intermediate-age
populations with ages of at least several 100 Myr if not several Gyr.
Furthermore, PNe have only been detected in two dSphs to date (Fornax
and Sagittarius).
Obviously one cannot expect to measure H\,{\sc ii} regions in the 
gas-deficient dSphs.

In order to compare not only {\em mean stellar metallicities} in
dIrrs and dSphs, but also the metallicities of the {\em same
populations} (i.e., of stars of similar age), we instead concentrated
on old Population II giants, which as mentioned earlier have been
detected in all LG dwarf galaxies.  We used (1) {\em old red giants} 
in dSphs and in the outskirts of dIrrs (where old populations dominate), 
(2) {\em spectroscopic abundances} wherever available (from own Keck
LRIS measurements and literature data from studies conducted at ESO,
NOAO, and Keck), and (3) {\em photometric abundances} elsewhere
(from comparison with globular cluster fiducials applied to our
own deep HST data from a WFPC2 snapshot survey and archival or 
literature data).  While the degree of homogeneity of the resulting
data set is not ideal, it is the best currently available one and
entirely based on well-calibrated empirical indicators.  

The resulting metallicity-luminosity relationship shows that even
when confined to old populations, there is a considerable offset 
between dSphs and dIrrs (Grebel et al.\ 2003).  {\bf At the same
galaxy luminosity, the old populations of dSphs are more metal-rich
than those of dIrrs}.  This indicates that in contrast to dIrrs,
dSphs must have experienced fairly rapid early enrichment.  
Together with various other factors, {\bf these evolutionary differences
make normal dIrrs unlikely progenitors of dSphs} (see also Binggeli
1994).  DIrr/dSph transition-type
galaxies, on the other hand, seem fairly plausible progenitors as 
explained in more detail in Grebel et al.\ (2003).

\subsection{Detailed abundance ratios}

Mean spectroscopic metallicities of individual stars provide crucial 
constraints on otherwise photometrically derived star formation histories
and allow one to break the age-metallicity degeneracy -- 
one of the prime areas of study with the new large telescopes. 

Another very important area for large telescopes is high-resolution
spectroscopy to measure individual abundance ratios.  In particular,
the determination of Fe, $\alpha$-, r-, and s-process element abundance 
ratios makes it possible to measure the modes and rates of star 
formation spectroscopically:
the relative contribution of supernovae of Type II vs.\ Ia, and that of
AGB stars, at different times during the evolution of the target galaxy.  
Importantly, this research permits one to compare the
abundance ratios of different types of dwarf galaxies with those
measured in various Galactic components and to constrain the building
block scenario from the chemical point of view. 

The [$\alpha$/Fe] ratios in dSphs (and dIrrs) at a given [Fe/H] are lower
than those measured in the Galactic halo, indicating either low star
formation rates in the dwarfs, loss of metals, or a larger contribution
from supernovae of Type Ia.  This is {\bf strong evidence against present-day
dSphs as the dominant contributors to the build-up of the Galactic halo}
(Shetrone et al.\ 2001).  For a discussion of the implications of recent
results for nucleosynthesis and galaxy evolution, see also Venn et al.\
(2004).  Extending these kinds of measurements and adding kinematic 
data as well is rapidly becoming one
of the major research areas for the world's largest telescopes such as
SALT, nicely
complemented by ongoing and future space missions such as HST, JWST, and
Gaia.

%


\begin{theacknowledgments}
I would like to thank Joanna Mikolajewska for organizing a very 
stimulating conference and for her patience while this contribution
was finished.
\end{theacknowledgments}


\bibliographystyle{aipprocl} 

\begin{thebibliography}{9}

\bibitem{Armand91}
T.~E.\ Armandroff, and G.~S.\ Da Costa, \emph{AJ}, \textbf{101}, 
1329--1337

\bibitem{Armand98}
T.~E.\ Armandroff, J.~E.\ Davies, and G.~H.\ Jacoby, 
\emph{AJ}, \textbf{116}, 2287--2296 (1998).

\bibitem{Armand99}
T.~E.\ Armandroff, G.~H.\ Jacoby, and J.~E.\ Davies, \emph{AJ}, \textbf{118}, 
1220--1229 (1999).

\bibitem{Bingg94}
B.\ Binggeli, ``Dwarf galaxies: a morphological overview'', in 
\emph{Panchromatic View of Galaxies. Their Evolutionary Puzzle}, edited by
G.\ Hensler, C.\ Theis, \& J.~S.\ Gallagher, Editions Frontiers,
Gif-sur-Yvette, 1994, pp.\ 173--191

\bibitem{Bouch03}
A.\ Bouchard, C.\ Carignan, and S. Mashchenko, \emph{AJ}, \textbf{126},
1295--1304 (2003).

\bibitem{Brown03}
T.~M.\ Brown, H.~C.\ Ferguson, E.\ Smith, R.~A.\ Kimble, A.~V.\ Sweigart,
A.\ Renzini, R.~M.\ Rich, and D.~A.\ VandenBerg, \emph{ApJ}, \textbf{592},
L17--L20 (2003).

\bibitem{Brown04}
T.~M.\ Brown, H.~C.\ Ferguson, E.\ Smith, R.~A.\ Kimble, A.~V.\ Sweigart,
A.\ Renzini, and  R.~M.\ Rich, \emph{AJ}, \textbf{127}, 2738--2752 (2004).

\bibitem{Bruens05}
C.\ Br\"uns, et al., \emph{A\&A}, \textbf{432}, 45--67 (2005).

\bibitem{Choi02}
P.~I.\ Choi, P.\ Guhathakurta, and K.~V.\ Johnston, \emph{AJ}, \textbf{124},
310--331 (2002).

\bibitem{Cole04}
A.~A.\ Cole, T.~A.\ Smecker-Hane, E.\ Tolstoy, T.~L.\ Bosler, and J.~S.
Gallagher, \emph{MNRAS}, \textbf{347}, 367--379 (2004).

\bibitem{Dekel03}
A.\ Dekel, and J.\ Woo, \emph{MNRAS}, \textbf{344}, 1131--1144 (2003).

\bibitem{Dolphin97}
A.~E.\ Dolphin, \emph{NewA}, \textbf{2}, 397--409 (1997).

\bibitem{Ferg02} 
A.~M.~N.\ Ferguson, M.~J.\ Irwin, R.~A.\ Ibata, G.~F.\ Lewis, and N.\ Tanvir, 
\emph{AJ}, \textbf{124}, 1452--1463 (2002).

\bibitem{Ferr00}
A.\ Ferrara, and E.\ Tolstoy, \emph{MNRAS}, \textbf{313}, 291--309 (2000).

\bibitem{Free02}
K.\ Freeman, and J.\ Bland-Hawthorn, \emph{ARA\&A}, \textbf{40}, 487--537 
(2002).

\bibitem{Gallagher03}
J.~S.\ Gallagher, G.~J.\ Madsen, R.~J.\ Reynolds, E.~K.\ Grebel, \&
T.~A.\ Smecker-Hane, \emph{ApJ}, \textbf{588}, 326--330 (2003).

\bibitem{Gallart96}
C.\ Gallart, A.\ Aparicio, G.\ Bertelli, and C.\ Chiosi,  \emph{AJ},
\textbf{112}, 1950--1968 (1996).

\bibitem{Gilmore02}
G.\ Gilmore, R.~F.~G.\ Wyse, and J.~E.\ Norris,  \emph{ApJ}, \textbf{574},
L39--L42 (2002).

\bibitem{Grebel97} 
E.~K.\ Grebel, \emph{Reviews in Modern Astronomy}, \textbf{10}, 27--59 (1997)

\bibitem{Grebel99} 
E.~K.\ Grebel, ``Evolutionary Histories of Dwarf Galaxies in the Local 
Group'', in \emph{The Stellar Content of the Local Group}, edited by
P.\ Whitelock \& R.\ Cannon, IAU Symp.\ 192, Astronomical Society of the
Pacific, San Francisco, 1999, pp.\ 17--38

\bibitem{Grebel00} 
E.~K.\ Grebel, ``The Star Formation History of the Local Group'', in
\emph{Star formation from the small to the large scale}, 
edited by F.\ Favata, A.A.\ Kaas, \& A.\ Wilson, 33rd ESLAB Symposium, 
ESA-SP 445, ESA, Noordwijk, 2000, pp.\ 87--98

\bibitem{Grebel01} 
E.~K.\ Grebel, \emph{A\&SSS}, \textbf{277}, 231--239 (2001).

\bibitem{Grebel04}
E.~K.\ Grebel, ``The Evolutionary History of Local Group Irregular Galaxies'',
in \emph{Origin and Evolution of the Elements}, edited by A.\ McWilliam \&
M.\ Rauch, Carnegie Observatories Astrophysics Series, Vol.\ 4, Cambridge
University Press, Cambridge, 2004, pp.\ 237--257 

\bibitem{GG99}
E.~K.\ Grebel, and P.\ Guhathakurta, \emph{ApJ}, \textbf{511}, L101--L105 
(1999).

\bibitem{GS99}
E.~K.\ Grebel, and P.~B.\ Stetson, ``Ground-based and WFPC2 Imaging of
Fornax: Spatial Variations in Star Formation History'', 
in \emph{The Stellar Content of the Local Group}, edited by
P.\ Whitelock \& R.\ Cannon, IAU Symp.\ 192, Astronomical Society of the
Pacific, San Francisco, 1999, pp.\ 165--169

\bibitem{GG04}
E.~K.\ Grebel, and J.~S.\ Gallagher, \emph{ApJ}, \textbf{610}, L89--L92 (2004).

\bibitem{GGH03}
E.~K.\ Grebel, J.~S.\ Gallagher, and D.\ Harbeck, \emph{AJ}, \textbf{125},
1926--1939 (2003).

\bibitem{Grill98}
C.~J.\ Grillmair, et al., \emph{AJ}, \textbf{115}, 144--151 (1998).

\bibitem{Harb01} 
D.\ Harbeck, et al., \emph{AJ}, \textbf{122}, 3092--3105 (2001).

\bibitem{Harb04}
D.\ Harbeck, J.~S.\ Gallagher, and E.~K.\ Grebel, \emph{AJ}, \textbf{127},
2711--2722 (2004).

\bibitem{Harb05}
D.\ Harbeck, J.~S.\ Gallagher,  E.~K.\ Grebel, A.\ Koch, and D.~B. Zucker,
\emph{ApJ}, in press (2005).

\bibitem{Holtz97}
J.~A.\ Holtzman, et al., \emph{AJ}, \textbf{113}, 656--668 (1997).

\bibitem{Hunter97}
D.\ Hunter, \emph{PASP}, \textbf{109}, 937--950 (1997).

\bibitem{Ibata94}
R.\ Ibata, G.\ Gilmore, and M.~J.\ Irwin, \emph{Nature}, \textbf{370},
194 (1994).

\bibitem{Ibata01} 
R.\ Ibata, M.\ Irwin, G.\ Lewis, A.~M.~N.\ Ferguson, and N.\ Tanvir, 
\emph{Nature}, \textbf{412}, 49--52 (2001).

\bibitem{Ikuta02}
C.\ Ikuta, and N.\ Arimoto, \emph{A\&A}, \textbf{391}, 55--65 (2002).

\bibitem{Jerjen00} 
H.\ Jerjen, B.\ Binggeli, and  K.~C. Freeman, \emph{AJ}, \textbf{119}, 
593--608 (2000).

\bibitem{KarKar99}
I.~D. Karachentsev, and V.~E.\ Karachentseva, \emph{A\&A}, \textbf{341}, 
355--356 (1999).

\bibitem{KKSG00} 
I.~D.\ Karachentsev, V.~E.\ Karachentseva, A.~A.\ Suchkov, and E.~K.\ Grebel, 
\emph{A\&AS}, \textbf{145}, 415--423 (2000).

\bibitem{Kar02a}
I.~D.\ Karachentsev, et al., \emph{A\&A}, \textbf{389}, 812--824 (2002a).

\bibitem{Klypin99} 
A.\ Klypin, A.~V.\ Kravtsov, O.\ Valenzuela, and F.\ Prada, \emph{ApJ}, 
\textbf{522}, 82--92 (1999).

\bibitem{Lejeune02}
T.\ Lejeune, and J.\ Fernandes, \emph{Observed HR Diagrams and Stellar
Evolution}, edited by T.\ Lejeune \& J.\ Fernandes, ASP Conf.\ Proc.,
Vol.\ 274, Astronomical Society of the Pacific, San Francisco, 2002

\bibitem{Mayer01}
L.\ Mayer, F.\ Governato, M.\ Colpi, B.\ Moore, T.\ Quinn, J.\ Wadsley,
J.\ Stadel, and G.\ Lake, \emph{ApJ}, \textbf{547}, L123--L127 (2001).

\bibitem{Martin04a} 
N.~F.\ Martin, N.~F., R.~A.\ Ibata, M.\ Bellazzini, M.~J.\ Irwin, 
G.~F.\ Lewis, and W.\ Dehnen, \emph{MNRAS}, \textbf{348}, 12--23 (2004a).

\bibitem{Martin04b}
N.~F.\ Martin, N.~F., R.~A.\ Ibata, B.~C.\ Conn, G.~F.\ Lewis, M.\ Bellazzini,
M.~J.\ Irwin, and A.~W.\ McConnachie, \emph{MNRAS}, \textbf{355}, L33--L37,
(2004b).

\bibitem{Mastro05}
C.\ Mastropietro, B.\ Moore, L.\ Mayer, J.\ Wadsley, and J.\ Stadel,
\emph{MNRAS}, submitted (astro-ph/0412312), (2005).

\bibitem{Mateo98} 
M.\ Mateo, \emph{ARA\&A}, \textbf{36}, 435--506 (1998).

\bibitem{Momany04}
Y.\ Momany, S.~R.\ Zaggia, P.\ Bonifacio, G.\ Piotto, F.\ De Angeli, 
L.~R.\ Bedin, and G.\ Carraro, \emph{A\&A}, \textbf{421}, L29--L32 (2004).

\bibitem{Moore99} 
B.\ Moore, S.\ Ghigna, F.\ Governato, G.\ Lake, T.\ Quinn, J.\ Stadel, and 
P.\ Tozzi, \emph{ApJ}, \textbf{524}, L19--L22 (1999).

\bibitem{Morr03} 
H.~L.\ Morrison, P.\ Harding, D.\ Hurley-Keller, and G.\ Jacoby, 
\emph{ApJ}, \textbf{596}, L183--L186 (2003).

\bibitem{Newberg02} 
H.~J.\ Newberg, et al., \emph{ApJ}, \textbf{569}, 245--274 (2002).

\bibitem{Odenk01}
M.\ Odenkirchen, et al., \emph{ApJ}, \textbf{548}, L165--L169 (2001).

\bibitem{Odenk03}
M.\ Odenkirchen, et al., \emph{AJ}, \textbf{126}, 2385--2407 (2003).

\bibitem{Palma03} 
C.\ Palma,  S.~R.\ Majewski, M.~H.\ Siegel, R.~J.\ Patterson, J.~C.\
Ostheimer, and R.\ Link, \emph{AJ}, \textbf{125}, 1352--1372 (2003).

\bibitem{Ped02} 
S.\ Pedraz, J.\ Gorgas, N.\ Cardiel, P.\ S\'anchez-Bl\'azquez, and R.\
Guzm\'an, \emph{MNRAS}, \textbf{322}, L59--L63 (2002).

\bibitem{Putman98}
M.~E.\ Putman, et al., \emph{Nature}, \textbf{394}, 752--754 (1998).

\bibitem{Richer98}
M.\ Richer, M.~L.\ McCall, \& G.\ Stasinska, \emph{A\&A}, \textbf{340},
67--76 (1998).

\bibitem{Rocha04}
H.~J.\ Rocha-Pinto, S.~R.\ Majewski, M.~F.\ Skrutskie, J.~D.\ Crane,
and R.~J.\ Patterson, \emph{ApJ}, \textbf{615}, 732--737 (2004).

\bibitem{Rutledge97}
G.~A.\ Rutledge, J.~E.\ Hesser, and P.~B.\ Stetson, \emph{PASP}, \textbf{109},
907--919 (1997).

\bibitem{Sersic68} 
J.~L.\ S\'ersic, \emph{Atlas de Galaxias Australes}, Obs.\ Astron.\ de 
C\'ordoba, C\'ordoba, 1968.

\bibitem{Shet01}
M.~D.\ Shetrone, P.\ C\^ot\'e, and W.~L.~W.\ Sargent, \emph{ApJ},
\textbf{548}, 592--608 (2001).

\bibitem{Smeck94}
T.~A.\ Smecker-Hane, P.~B.\ Stetson, J.~E.\ Hesser, and M.~D.\ Lehnert,
\emph{AJ}, \textbf{108}, 507--513 (1994)

\bibitem{Susa04}
H.\ Susa, and M.\ Umemura, \emph{ApJ}, \textbf{610}, L5--L8 (2004).

\bibitem{Tassis03} 
K.\ Tassis, T.\ Abel, G.~L.\ Bryan, and M.~L.\ Norman, \emph{ApJ}, 
\textbf{587}, 13--24 (2003).

\bibitem{Tolstoy96}
E.\ Tolstoy, and A.\ Saha, \emph{ApJ}, \textbf{462}, 672--683 (1996).

\bibitem{Tosi91}
M.\ Tosi, L.\ Greggio, G.\ Marconi, and P.\ Focardi, \emph{AJ}, \textbf{102},
951--974 (1991).

\bibitem{vdB94}
S.\ van den Bergh, \emph{ApJ}, \textbf{428}, 617--619 (1999).

\bibitem{vdB99} 
S.\ van den Bergh, \emph{A\&ARv}, \textbf{9}, 273--318 (1999).

\bibitem{vdB00} 
S.\ van den Bergh, \emph{The Galaxies of the Local Group}, 
Cambridge Astrophysics Series Vol.\ 35, Cambridge University Press,
Cambridge, 2000.

\bibitem{vdM01}
R.~P.\ van der Marel, \emph{AJ}, \textbf{122}, 1827--1843 (2001).

\bibitem{vLeeuw02}
F.\ van Leeuwen, J.~D.\ Hughes, and G.\ Piotto, \emph{$\omega$ Centauri:
A Unique Window into Astrophysics}, edited by F.\ van Leeuwen, J.~D.\ Hughes,
\& G.\ Piotto,  ASP Conf.\ Ser.\ 265, Astronomical Society of the Pacific,
San Francisco, 2002.

\bibitem{Venn04}
K.~A.\ Venn, M.\ Irwin, M.~D.\ Shetrone, C.~A.\ Tout, V.\ Hill, and E.\ 
Tolstoy, \emph{AJ}, \textbf{128}, 1177-1195 (2004).

\bibitem{Wilk04}
M.~I.\ Wilkinson, J.~T.\ Kleyna, N.~W.\ Evans, G.~F.\ Gilmore, M.~J.\
Irwin, and E.~K.\ Grebel, \emph{ApJ}, \textbf{611}, L21--L24 (2004).

\bibitem{WHI99} 
A.~B.\ Whiting, G.~K.~T.\ Hau, and M.\ Irwin, \emph{AJ}, \textbf{118}, 
2767--2774 (1999).

\bibitem{Whiting02}
A.~B.\ Whiting, G.~K.~T.\ Hau, and M.\ Irwin, \emph{ApJS}, \textbf{141}, 
123--146 (2002).

\bibitem{Wyse02}
R.~F.~G.\ Wyse, G.\ Gilmore, M.~L.\ Houdashelt, S.\ Feltzing, L.\ Hebb,
J.~S.\ Gallagher, and T.~A.\ Smecker-Hane, \emph{NewA}, \textbf{7},
395--433 (2002).

\bibitem{Yanny03}
B.\ Yanny, et al., \emph{ApJ}, \textbf{588}, 824--841 (2003).

\bibitem{Zar00}
D.\ Zaritsky, J.\ Harris, E.~K.\ Grebel, and I.~B.\ Thompson, \emph{ApJ}, 
\textbf{534}, L53--L56 (2000).

\bibitem{Zucker04a}
Zucker, D.~B., et al., \emph{ApJ}, \textbf{612}, L121--L124 (2004a).

\bibitem{Zucker04b}
Zucker, D.~B., et al., \emph{ApJ}, \textbf{612}, L117--L120 (2004b).


\end{thebibliography}



\end{document}